\documentclass[aps,prc,reprint,showpacs]{revtex4-1}

\usepackage{graphicx}
\usepackage{dcolumn}
\usepackage{bm}

\include{00README.XXX}

\begin{document}


\title{Transmission resonance spectroscopy in the third minimum of $^{232}$Pa}
\author{L. Csige$^{1,2}$}
\author{M. Csatl\'os$^2$}
\author{T. Faestermann$^3$}
\author{J. Guly\'as$^2$}
\author{D. Habs$^{1,4}$}
\author{R. Hertenberger$^1$}
\author{M. Hunyadi$^2$}
\author{A. Krasznahorkay$^2$}
\author{H. J. Maier$^1$}
\author{P. G. Thirolf$^1$}
\author{H.-F. Wirth$^3$}
\affiliation{
$^1$Ludwig-Maximilians-Universit\"at M\"unchen, D-85748 Garching, Germany \\
$^2$Institute of Nuclear Research of the Hungarian Academy of Sciences (ATOMKI), Post Office Box 51, 
H-4001 Debrecen, Hungary \\
$^3$Technische Universit\"at M\"unchen, D-85748 Garching, Germany \\
$^4$Max-Planck-Institute for Quantum Optics, D-85748 Garching, Germany}

\date{\today}

\begin{abstract}
The fission probability of $^{232}$Pa was measured as a function of the excitation energy in order to search for hyperdeformed (HD) transmission resonances using the ($d,pf$) transfer reaction on a radioactive $^{231}$Pa target. The experiment was performed at the Tandem accelerator of the Maier-Leibnitz Laboratory (MLL) at Garching using the $^{231}$Pa($d,pf$) reaction at a bombarding energy of $E_{\text{d}}$=12 MeV and with an energy resolution of $\Delta E$=5.5 keV. Two groups of transmission resonances have been observed at excitation energies of $E^*$=5.7 and 5.9 MeV. The fine structure of the resonance group at $E^*$=5.7 MeV could be interpreted as overlapping rotational bands with a rotational parameter characteristic to a HD nuclear shape ($\hbar^2/2\Theta$=2.10$\pm$0.15 keV). The fission barrier parameters of $^{232}$Pa have been determined by fitting TALYS 1.2 nuclear reaction code calculations to the overall structure of the fission probability. From the average level spacing of the $J$=4 states, the excitation energy of the ground state of the 3$^{\text{rd}}$ minimum has been deduced to be $E_{\text{III}}$=5.05$^{+0.40}_{-0.10}$ MeV.
\end{abstract}


\maketitle

The observation of discrete $\gamma$ transitions between hyperdeformed (HD) nuclear states represents one of the last frontiers of high-spin physics. Although a large community with 4$\pi$ $\gamma$ arrays was searching for HD states in very long experiments, no discrete HD $\gamma$ transition was found in the mass region of $A\approx$100-130 \cite{la95,la96,wi97,nya05,hers07}. On the other hand, the existence of low-spin hyperdeformation in the third minimum of the fission barrier is established both experimentally and theoretically in the actinide region \cite{th02,kr11}. Observing transmission resonances as a function of the excitation energy caused by resonant tunneling through excited states in the third minimum of the potential barrier can specify the excitation energies of the HD states. Moreover, the observed states could be ordered into rotational bands and the moments of inertia of these bands can characterize the underlying nuclear shape, proving that these states have indeed a HD configuration.

Regarding hyperdeformation, the double-odd nucleus $^{232}$Pa is of great interest. Even though low-spin hyperdeformation has already proved to be a general feature of uranium \cite{csi09,csa05,kr99} and thorium isotopes \cite{bl88}, no HD state has been found in protactinium isotopes so far. In particular, the level scheme of the odd-odd $^{232}$Pa is completely unknown in the 1$^{\text{st}}$ minimum of the potential barrier, only the ground-state properties are known at present (I$_{\text{gs}}^{\pi}$=2$^-$) \cite{br06}. The fine structure of the fission resonances of $^{232}$Pa has been studied so far only via the ($n,f$) reaction \cite{pl81} with high resolution, but the results showed no convincing evidence on the existence of HD states. A possible reason was the rather limited momentum transfer of the ($n,f$) reaction at that low neutron energy ($E_{\text{n}}\approx$100 keV), which did not allow for the population of rotational bands. In contrast, the ($d,p$) reaction can transfer considerable angular momentum, thus full sequences of rotational states with higher spins can be excited. The experimental ($n,f$) cross-section was used very recently to deduce the fission barrier parameters of $^{232}$Pa by performing cross-section calculations with the EMPIRE 2.19 nuclear reaction code \cite{herm07}, in which the optical model for fission was extended to treat double- and triple-humped fission barriers. The fission barrier parameters of  $^{232}$Pa were determined to be $E_{\text{A}}$=5.92, $E_{\text{BI}}$=6.3 and $E_{\text{BII}}$=6.34 MeV \cite{si06}. This result suggested to expect the appearance of HD resonances in the excitation energy region between $E^*$=5.9 and $E^*$=6.3 MeV.

In our experiment, the fission probability of $^{232}$Pa was measured as a function of the excitation energy with high resolution in order to search for HD rotational bands using the $^{231}$Pa($d,pf$) reaction. The experiment was carried out at the Tandem accelerator of the Maier-Leibnitz-Laboratory (MLL) at Garching employing the $^{231}$Pa($d,pf$) reaction with a bombarding energy of $E_{\text{n}}$=12 MeV to investigate the fission probability of $^{232}$Pa in the excitation energy region of $E^*$=5.5-6.2 MeV. Enriched (99\%), 70 $\mu$g/cm$^2$ thick radioactive target of $^{231}$Pa was used on a 20 $\mu$g/cm$^2$ thick carbon backing. The ground-state $Q$-value for the reaction is $Q$=3.324 MeV, which was calculated using the NNDC $Q$-value calculator. The excitation energy of the fissioning nucleus was derived from the kinetic energy of the outgoing protons, that was measured by the Garching Q3D magnetic spectrograph \cite{en79} set at $\Theta_{\text{lab}}$=139.4$^{\circ}$ relative to the beam direction. The well-known lines of the $^{208}$Pb($d,p$) reaction were applied to perform the energy calibration of the focal plane detector \cite{wi01}. The experimental energy resolution was deduced to be $\Delta E$=5.5 keV (FWHM) in the energy region of our interest. Fission fragments were detected in coincidence with the outgoing protons by two position-sensitive avalanche detectors (PSADs) with a solid angle coverage of 20\% of 4$\pi$.

\begin{figure}
\centering
\includegraphics[width=8.5cm]{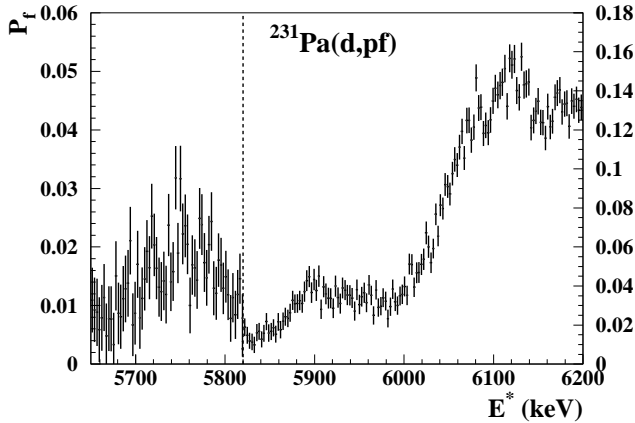}
\caption{\label{fig:fisprob} The measured fission probability of $^{232}$Pa in the excitation energy range of $E^*$=5.5-6.2 MeV. Two resonance groups have been observed around $E^*$=5.75 and 5.9 MeV, respectively, in agreement with the results of a previous ($n,f$) experiment \cite{pl81}. Below $E^*$=5.82 MeV a magnified scale of the y-axis was used for better visibility of the resonance structure.}
\end{figure}

The measured high-resolution fission probability spectrum of $^{232}$Pa is shown in Fig.~\ref{fig:fisprob} as a function of the excitation energy of the fissioning nucleus in the region of $E^*$=5.5-6.2 MeV. The random coincidence contribution was subtracted by using the well-defined flight time difference of protons and fission fragments. Two resonance groups can be clearly seen at $E^*$=5.75 and 5.9 MeV in a fair agreement with the results of a previous ($n,f$) experiment \cite{pl81}. Below $E^*$=5.82 MeV a magnified scale of the y-axis was used for better visibility of the resonance structure. In the ($n,f$) experiment, low-energy neutrons ($E_{\text{n}}$=120-420 keV) were used to populate the states in the compound nucleus. In this case s-wave neutron capture is the dominant process and the transfer momentum is principally limited to 1$\hbar$, thus rotational bands cannot be excited. On the other hand, the fission fragment angular distribution (FFAD) data supported a $K$=3$^+$ assignment for the resonance at $E_{\text{n}}$=156.7 keV. Together with a possible $K$=3$^-$ assignment for the resonance at $E_{\text{n}}$=173.3 keV, which could not be ruled out by the FFAD data, these two resonances could be the bandheads of two close-lying $K$-bands with opposite parities, a well-known consequence of the octupole deformation in the HD minimum of the fission barrier. However, having no information on the moment of inertia, this result could not be considered as a clear evidence on the existence of a HD minimum as also stated by the authors.

\begin{figure}
\centering
\includegraphics[width=8.5cm]{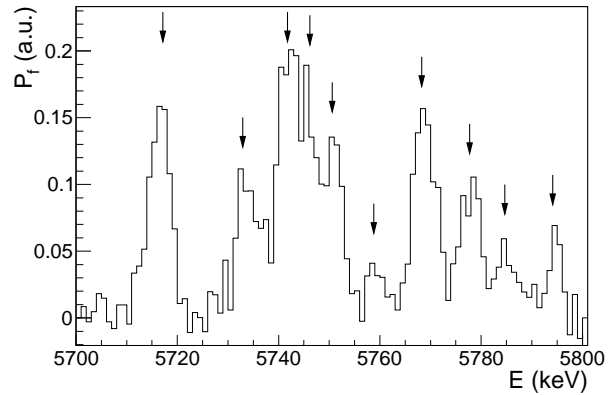}
\caption{\label{fig:fit} The result of the Markov-chain peak searching algorithm applied to the histogram, which is shown in Fig.~\ref{fig:fisprob} ($E^*$=5.7-5.8 MeV). The continuous fission background stemming from the non-resonant tunneling process is subtracted. The positions of the identified resonances are indicated by arrows.}
\end{figure}

Due to the low neutron separation energy of $^{232}$Pa ($S_{\text{n}}$=5.455 MeV), the fission probability is rather small, which resulted in a very limited statistics at deep sub-barrier energies. Therefore, to suppress the statistical fluctuations of the excitation energy spectrum and to identify the resonances unequivocally, we applied a widely used peak-searching method, the so-called Markov-chain algorithm \cite{si96} to the data. This method can also be used to subtract the continuous, exponentially rising fission background originating from the non-resonant tunneling process through the fission barrier. In the generated spectrum, a number of sharp resonances could be clearly identified between $E^*$=5.7 and 5.8 MeV as indicated by arrows in Fig.~\ref{fig:fit}.

The limited statistics of the experiment did not allow to extract the spin information from the angular data. However, based on the results of the ($n,f$) experiment \cite{pl81}, the first resonance around $E^*$=5.72 MeV can be identified as the bandhead of a $K$=3 band with more members in the present experiment owing to the larger transfer momentum of the ($d,p$) reaction. The resonance group at $E^*$=5.9 MeV could not be resolved into individual resonances as a consequence of the large level density (thus strongly overlapping) at this high excitation energy.

\begin{figure}
\centering
\includegraphics[width=8.5cm]{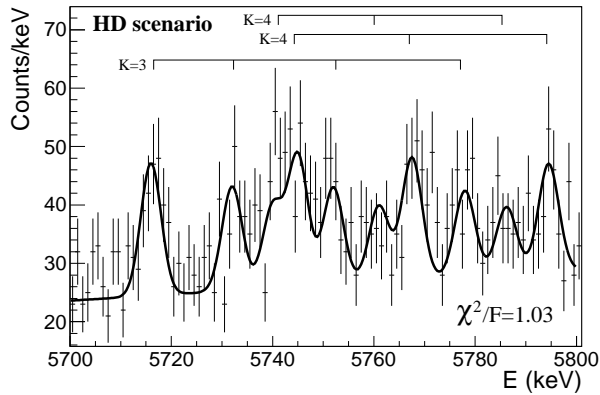}
\caption{\label{fig:fiterrora} Excitation energy spectrum of $^{232}$Pa with statistical errors. The result of the fitting procedure with HD rotational bands is indicated by the solid line. The picket fence structure of the rotational bands together with the $K$ values of the bands are also shown. The quality of the fit (the reduced $\chi^2$ value) is $\chi^2/F$=1.03.}
\end{figure}

To allow for an identification of the underlying structure as either resulting from a hyperdeformed (HD) or superdeformed (SD) configuration, the observed resonances have been fitted with overlapping rotational bands assuming both scenarios. Gaussians were used to describe the different band members with a width fixed to the experimental resolution ($\Delta E$=5.5 keV). During the fitting procedure, the energy of the bandheads and the intensity of the band members were treated as free parameters and a common rotational parameter was adopted for each band ($\hbar^2/2\Theta$=2.10$\pm$0.15 keV for the HD scenario). Since the population of the different spins varies only slightly with excitation energy, the intensity ratios of the band members were kept to be constant. The result of the fitting procedure is presented in Fig.~\ref{fig:fiterrora}. The picket fence structure of the three rotational bands is indicated in the figure as well as the quality of the fit ($\chi^2/F$=1.03 with $F$=99).

\begin{figure}
\centering
\includegraphics[width=8.5cm]{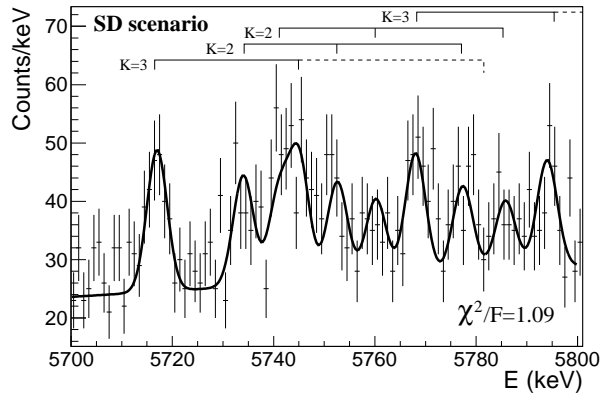}
\caption{\label{fig:fiterrorb} Excitation energy spectrum of $^{232}$Pa with the result of the fitting procedure (solid line) assuming SD rotational bands. The picket fence structure and the $K$ values of the bands are also indicated. The dashed lines represent the missing members of the bands. The quality of the fit (the reduced $\chi^2$ value) is $\chi^2/F$=1.09.}
\end{figure}

Given the resonance positions determined by the Markov-chain algorithm, we also tested the assumption of an underlying SD rotational band configuration generating the observed resonance structure. In this scenario (Fig.~\ref{fig:fiterrorb}), our data could be described by four rotational bands with $K$ value assignments of $K$=3,2,2 and 3, respectively. In this case, the rotational parameter of the bands was $\hbar^2/2\Theta$=3.3$\pm$0.2 keV, which is characteristic to SD nuclear shapes. The quality of the fit is $\chi^2/F$=1.09 ($F$=99). However, there are several arguments disfavoring this interpretation. As one can see in Fig.~\ref{fig:fiterrorb}, only two resonances could be combined to form the first SD rotational band (K=3), while at expected positions of further members no resonances were observed. Moreover, we have no proof on the existence of a third member of the last $K$=3 band (expected at $E^*$=5800 MeV) due to the high level density in the second resonance structure around $E^*$=5.9 MeV. The FFAD data \cite{pl81} also disagree with the assignment of $K$=2 for the second and third rotational band. Furthermore, the level density should be much higher in the deep second minimum ($E_{\text{II}}$=1.9 MeV according to Ref.~\cite{si06}) at this high excitation energy ($E$=5.7 MeV).

As a conclusion, the observed resonance fine structure can most convincingly be described as a sequence of three overlapping rotational bands with assignments of $K$=3,4 and 4 for the bandheads at $E^*$=5717, 5740 and 5745 keV, respectively, and with a rotational parameter ($\hbar^2/2\Theta$=2.10$\pm$0.15 keV) characteristic for HD nuclear shapes.

The depth of the third minimum was determined by comparing the experimentally obtained average level spacings of the $J$=4 members ($D_{J=4}$=9 keV) of the HD rotational bands (Fig.~\ref{fig:fiterrora}) with the calculated ones using the back-shifted Fermi gas (BSFG) description of the level density. The level density of a given nucleus has usually been determined by adjusting the level density parameters to obtain the best description of the low-energy cumulative discrete level schemes as well as the s-wave neutron resonance spacings. However, in the case of $^{232}$Pa no level scheme is available as already pointed out, so we could not extract the NLD parameters this way. On the other hand, systematic investigations of the NLD showed, that very simple analytic expressions can be used to estimate the NLD parameters involving some basic nuclear quantities like the shell correction energy and the deuteron pairing energy \cite{eg09}. Following the concept of Ref.~\cite{eg12}, the back-shifted Fermi gas (BSFG) level density parameters of $^{232}$Pa were estimated to be $a$=23.55 MeV$^{-1}$ and $E_1$=-1.103 MeV.

In order to deduce the excitation energy $E_{\text{III}}$ of the ground state in the third minimum, an excitation energy of $U_{\text{III}}$=$E-E_1-E_{\text{III}}$ was used in the formulas of Ref.~\cite{eg12}, where $E_1$ and $E_{\text{III}}$ stand for the energy backshift of the 1$^{\text{st}}$ and the 3$^{\text{rd}}$ potential minimum (with respect to the 1$^{\text{st}}$ minimum), respectively.

\begin{figure}
\centering
\includegraphics[width=8.5cm]{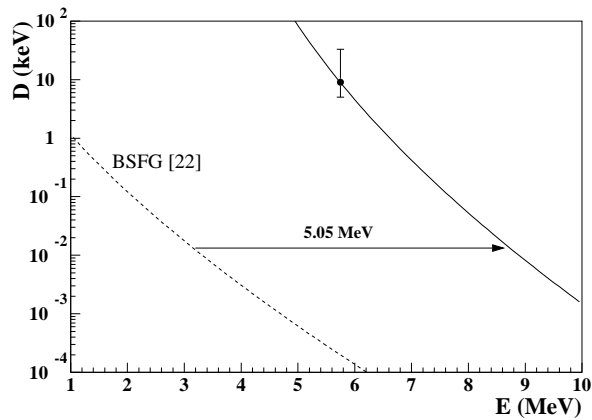}
\caption{\label{fig:ld} Experimental and calculated average level spacings of the $J$=4 states in the function of the excitation energy for $^{232}$Pa. Calculated level spacings are indicated by a solid and a dashed line for representing the values in the first and in the third potential minimum (with $E_{\text{III}}$=5.05 MeV), respectively. Within the calculation of the BSFG level densities, we used the same parametrization as in Ref.~\cite{eg09}. The error bar of the experimental point (circle) represents the upper and lower limits of the observed $J$=4 level spacings.}
\end{figure}

To match the calculated level spacings to our experimental point (circle in Fig.~\ref{fig:ld}) the value of $E_{\text{III}}$ was varied. We obtained the best description with $E_{\text{III}}$=5.05 MeV for the excitation energy of the ground state in the third minimum as indicated by solid line in Fig.~\ref{fig:ld}. To estimate the uncertainty of $E_{\text{III}}$, we used an upper and a lower limit for the average level spacings. The smallest observed experimental spacing ($D$=5 keV) was taken as the lower limit, while the upper limit was chosen to be $D$=24 keV by assuming three equally distributed $J$=4 states in the excitation energy range of $E^*$=5.7-5.8 MeV. These limits are indicated in Fig.~\ref{fig:ld} as asymmetric error bars of the experimental point. Our final result is $E_{\text{III}}$=5.05$^{+0.40}_{-0.10}$ MeV, which indicates a less deep minimum for $^{232}$Pa in contrast to our previous results on the even-even uranium isotopes \cite{kr11}, while, however, still being significantly deeper than claimed in a recent theoretical study in this mass range \cite{ko12b}. On the other hand, our present result is in a good agreement with the result of Ref.~\cite{si06}, where the third minimum was found to be $E_{\text{III}}$=5.4 MeV.

To extract the fission barrier parameters of $^{232}$Pa, we performed cross-section calculations on the $^{231}$Pa($d,pf$) reaction using the TALYS 1.2 nuclear reaction code \cite{ko05}, the only available code that can calculate exclusive fission cross-sections with particle spectra for transfer reactions. In the code, the fission transmission coefficients are calculated following the concept of the Hill-Wheeler formalism, which then enter the Hauser-Feshback statistical model to compete with the particle and photon emission. The fission barrier is parametrized by smoothly joint parabolas, and the barrier parameters, namely the heights ($E_{\text{A,B1,B2}}$) and curvature energies ($\hbar\omega_{\text{A,B1,B2}}$) of a triple-humped fission barrier, are given as input parameters. 

A very important ingredient of the cross-section calculations is the nuclear level density (NLD), both at the equilibrium deformation and at the saddle points. In contrast to the level densities at normal deformation, the saddle level densities generally suffer from a serious lack of experimental information, however, a good approximation can be obtained by introducing additional constants to the ground-state NLD to describe the rotational and vibrational enhancements at large deformations.

\begin{figure}
\centering
\includegraphics[width=8.5cm]{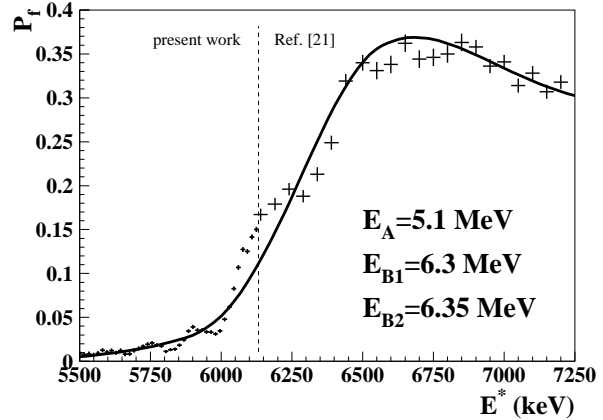}
\caption{\label{fig:talys} Experimental fission probability of $^{232}$Pa measured in the present experiment (below $E^*$=6.2 MeV) and in a previous, low-resolution measurement (above $E^*$=6.2 MeV) \cite{ba74} together with the result of the TALYS 1.2 calculation (continuous line).}
\end{figure}

\begin{figure}
\centering
\includegraphics[width=8.5cm]{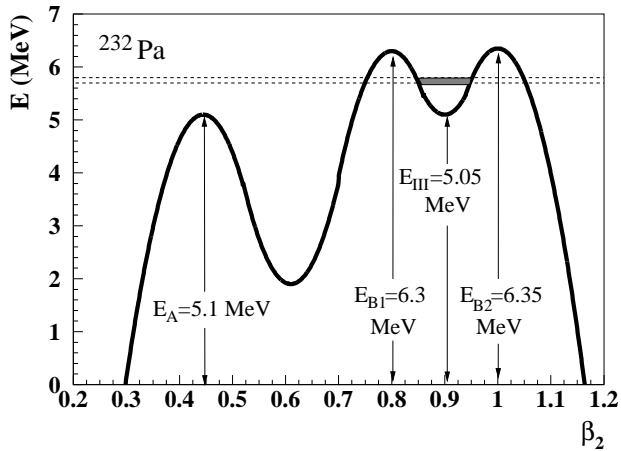}
\caption{\label{fig:pot} Triple-humped fission barrier of $^{232}$Pa as a result of the present study. The energy region of the observed HD resonances is indicated by horizontal dashed lines (and the hatched area). The curvature energies are $\hbar\omega_{\text{A,B1,B2}}$=1.0 MeV.}
\end{figure}

In Fig.~\ref{fig:talys} the experimental fission probability of $^{232}$Pa is shown in the excitation energy interval of $E^*=$5.2-7.25 MeV together with the result of the TALYS calculation (represented by a solid line) and the obtained fission barrier parameters. The data points of the present experiment ($E^*<$6.2 MeV) were extended by a result of a previous, low-resolution ($\Delta E$=55 keV) measurement \cite{ba74} to cover a larger energy range. Class-II (SD) and class-III (HD) states were not introduced into the calculations, so the resonance region could not be reproduced at low excitation energies, however, at this level we aimed at extracting the barrier parameters from the overall structure, the slope and the saturation of the fission probability. Nevertheless, our final parameter set is in good agreement with the results of Ref.~\cite{si06,si08}, where the EMPIRE 2.19 nuclear code was used to calculate the neutron-induced fission cross-section of $^{232}$Pa and fitted to the experimental cross-section. However, our calculation suggests a slightly lower inner barrier ($E_{\text{A}}$=5.1 MeV), taking into account also the relatively large uncertainty of the determination of the inner barrier height. Curvature energies of $\hbar\omega_{\text{A,B1,B2}}$=1.0 MeV have been used in the calculations. Our fission barrier parameters are consistent with the appearance of class-III resonances between $E^*$=5.7 and 5.8 MeV and disfavoring the SD interpretation of the resonances. Fig.~\ref{fig:pot} shows the triple-humped fission barrier of $^{232}$Pa as a result of the present study. The energy region of the observed HD resonances is indicated by two dashed lines.

\begin{figure}
\centering
\includegraphics[width=8.5cm]{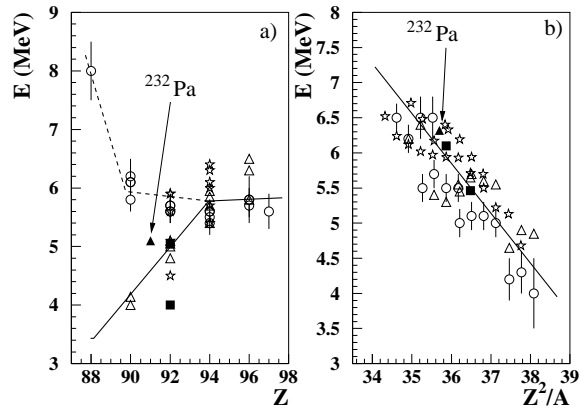}
\caption{\label{fig:sys} Calculated and experimental (a) inner ($E_{\text{A}}$) and (b) outer ($E_{\text{B}}$) barrier heights as a function of the atomic number ($Z$) and fissility parameter ($Z^2/A$), respectively, for actinide nuclei in the region of the ''island of fission isomers'' ($Z$=88-97). Clear tendencies can be seen for both barriers as illustrated by solid lines. The data points were taken from Ref.~\cite{bj80} (open circles), Ref.~\cite{csa05,csi09} (full squares),  Ref.~\cite{lu12} (open triangles) and Ref.~\cite{ko10,ko12} (open stars). Present results on $^{232}$Pa are also shown (full triangles). The dashed line represents the tendency of empirical inner barrier heights, which were determined by using the double-humped fission barrier concept \cite{bj80}.}
\end{figure}

Figure \ref{fig:sys} shows our present results on the inner [Fig.~\ref{fig:sys}(a), as a function of the nuclear charge] and outer [Fig.~\ref{fig:sys}(b), as a function of the fissility parameter] barrier heights of $^{232}$Pa together with the most recent experimental (empirical) and theoretical fission barrier parameters of even-even actinide nuclei in order to visualize systematic trends. The data points were taken from Ref.~\cite{bj80} (open circles), Ref.~\cite{csa05,csi09} (full squares),  Ref.~\cite{lu12} (open triangles) and Ref.~\cite{ko10,ko12} (open stars). For triple-humped barriers, the average of the two outer barriers ($<E_{\text{B1}},E_{\text{B2}}>$) is indicated. The data for the inner and outer barrier heights ($E_{\text{A}}$ and $E_{\text{B}}$) reveal clear trends as a function of the atomic number and fissility parameter, respectively, as illustrated by the two solid lines. Our new data points for $^{232}$Pa (full triangles) agree reasonably well with these observed tendencies. The dashed line in panel (a) shows the tendency of empirical inner barrier heights determined by using the double-humped fission barrier concept \cite{bj80}, which failed in predicting the most characteristic features of the fission cross-sections of the light actinides and gave rise to the well-known ''Thorium-anomaly'' problem, which was resolved by introducing the triple-humped fission barrier concept.

Summarizing our results, we measured the fission probability of $^{232}$Pa with high resolution using the $^{231}$Pa($d,pf$) transfer reaction to deduce the fission barrier parameters of $^{232}$Pa and search for hyperdeformed fission resonances. Sharp transmission resonances have been observed at excitation energies between $E^*$=5.7-5.8 MeV. These resonance structures could be interpreted most convincingly as overlapping rotational bands with a moment of inertia characteristic of hyperdeformed nuclear shapes ($\hbar^2/2\Theta$=2.10$\pm$0.15 keV). We found, for the first time, conclusive evidence on hyperdeformed configurations in protactinium isotopes and even more general in an odd-odd nucleus. The fission barrier parameters of $^{232}$Pa have been deduced by fitting calculated fission probability to the experimental values using the TALYS 1.2 nuclear reaction code. From the average level spacing of the observed resonances the excitation energy of the ground state of the 3$^{\text{rd}}$ was determined to be $E_{\text{III}}$=5.05$^{+0.40}_{-0.10}$ MeV corresponding to a depth of the third well of 1.25$^{+0.10}_{-0.40}$ MeV. The deduced fission barrier parameters agrees reasonably well with the results of Ref.~\cite{si06} and support our interpretation of the hyperdeformed fission resonances.

The work has been supported by DFG under HA 1101/12-2 and UNG 113/129/0, the DFG Cluster of Excellence ''Origin and Structure of the Universe'', and the Hungarian OTKA Foundation No. K72566.


\begin{thebibliography}{100}

\bibitem{la95} D.R. LaFosse et al., Phys. Rev. Lett. \textbf{74}, 5186 (1995).
\bibitem{la96} D.R. LaFosse et al., Phys. Rev. C \textbf{54}, 1585 (1996).
\bibitem{wi97} J.N. Wilson \textit{et al.},  Phys. Rev. C \textbf{56}, 2502 (1997).
\bibitem{nya05} B.M. Nyak\'o \textit{et al.}, Acta Phys. Pol. \textbf{B36}, 1033 (2005).
\bibitem{hers07} B. Herskind \textit{et al.}, Acta Phys. Pol. \textbf{B38}, 1421 (2007), and references therein.
\bibitem{th02} P.G. Thirolf and D. Habs, Prog. Part. Nucl. Phys. \textbf{49}, 325 (2002).
\bibitem{kr11} A. Krasznahorkay, in: Handbook of Nuclear Chemistry, Springer Verlag, 2011, p. 281.
\bibitem{csi09} L. Csige \textit{et al.}, Phys. Rev. C \textbf{80}, 011301 (2009).
\bibitem{csa05} M. Csatl\'os \textit{et al.}, Phys. Lett. \textbf{B615}, 175 (2005).
\bibitem{kr99} A. Krasznahorkay \textit{et al.},  Phys. Lett. \textbf{B461}, 15 (1999).
\bibitem{bl88} J. Blons et al., Nucl. Phys. \textbf{A477}, 231 (1988).
\bibitem{br06} E. Browne, Nuclear Data Sheets \textbf{107}, 2579 (2006).
\bibitem{pl81} S. Plattard \textit{et al.}, Phys. Rev. Lett. \textbf{46}, 633 (1981).
\bibitem{herm07} M. Herman \textit{et al.}, Nucl. Data Sheets, \textbf{108} 2655 (2007).
\bibitem{si06} M. Sin, R. Capote, A. Ventura, M. Herman and P. Oblozinsky, Phys. Rev. C \textbf{74}, 014608 (2006).
\bibitem{en79} H.A. Enge, Nucl. Inst. Meth. \textbf{162}, 161 (1979).
\bibitem{wi01} H.F. Wirth, Ph.D. thesis, Technische Universit\"at Munich, 2001.
\bibitem{si96} Z.K. Silagadze, NIM \textbf{A376} 451 (1996).
\bibitem{eg09} T. von Egidy and D. Bucurescu, Phys. Rev. C \textbf{80}, 054310 (2009).
\bibitem{eg12} T. von Egidy and D. Bucurescu, Journal of Physics: Conference Series \textbf{338}, 012028 (2012).
\bibitem{ko12b} M. Kowal and J. Skalski, arXiv:1203.4449v1 (2012).
\bibitem{ko05} A.J. Konig et al., AIP Conf. Proc. \textbf{769}, 1154 (2005).
\bibitem{ba74} B.B. Back et al., Phys. Rev. C \textbf{10}, 1948 (1974).
\bibitem{si08} M. Sin and R. Capote, Phys. Rev. C \textbf{77}, 054601 (2008).
\bibitem{bj80} S. Bj\o rnholm and J.E. Lynn, Rev. Mod Phys. \textbf{52}, 725 (1980).
\bibitem{lu12} B.-N. Lu, E.-G. Zhao and S.-G. Zhou, Phys. Rev. C \textbf{85}, 011301 (2012).
\bibitem{ko10} M. Kowal, P. Jachimowicz, and A. Sobiczewski, Phys. Rev. C \textbf{82}, 014303 (2010).
\bibitem{ko12} P. Jachimowicz, M. Kowal and J. Skalski, Phys. Rev. C \textbf{85}, 034305 (2012).

\end{thebibliography}
\end{document}